\documentclass[12pt]{iopart}

\usepackage{iopams}
\usepackage[dvipdf]{graphicx}
\usepackage{tabularx}
\usepackage{color} 
\usepackage{bm}

\begin{document}

\title[The equilibrium phase in heterogeneous Hertzian chains]{The equilibrium phase in heterogeneous Hertzian chains}

\author{Michelle Przedborski$^1$, Surajit Sen$^2$, Thad A. Harroun$^1$}
\address{$^1$ Department of Physics, Brock University, St. Catharines, Ontario, Canada}
\address{$^2$ Department of Physics, SUNY Buffalo, USA}
\ead{thad.harroun@brocku.ca}

\date{\today}

\begin{abstract}
We examine the long-term behaviour of non-integrable, energy-conserved, 1D systems of macroscopic grains interacting via a contact-only generalized Hertz potential and held between stationary walls. We previously showed that in homogeneous configurations of such systems, energy is equipartitioned at sufficiently long times, thus these systems ultimately reach thermal equilibrium. 
Here we expand on our previous work to show that heterogeneous configurations of grains also reach thermal equilibrium at sufficiently long times, as indicated by the calculated heat capacity. We investigate the transition to equilibrium in detail and introduce correlation functions that indicate the onset of the transition.
\end{abstract}

\maketitle

\section{Introduction}
In previous works~\cite{Przedborski2017a,Przedborski2017b}, we showed that 1D homogeneous systems of discrete macroscopic grains interacting via a power-law contact potential and held between fixed walls thermalizes an initial solitary wave (SW), ultimately transitioning to an equilibrium phase. The properties of this phase are well-predicted by treating the chain at long times as a 1D gas of interacting spheres in a microcanonical ($\bm{NVE}$) ensemble~\cite{Scalas2015}. In the present manuscript, we look more closely at the transition to equilibrium by using statistical tests to show that the long-term dynamics is ergodic, examine the behaviour of various correlation functions close to the onset of the transition, and extend the analysis to heterogeneous systems.

The discrete, non-integrable systems under consideration have been the focus of a number of recent investigations ~\cite{Nesterenko1983,Nesterenko1985,Nesterenko1995,Sinkovits1995,Sen1996,Coste1997,Sen1998,Chatterjee1999,Hinch1999,Hong1999,Ji1999,Manciu1999,Hascoet2000,Manciu2000,
Sen2001,Nesterenko2001,Sen2001b,Manciu2002,
Rosas2003,Nakagawa2003,
Rosas2004,Sen2004,
Nesterenko2005,Sokolow2005,Hong2005,Mohan2005,Sen2005,Job2005,
Doney2006,Melo2006,Daraio2006,
Job2007,Sokolow2007,Zhen2007,Avalos2007,
Sen2008,
Herbold2009,Doney2009,Job2009,Theocharis2009,
Boechler2010,Theocharis2010,
Santibanez2011,Breindel2011,Avalos2011,
Takato2012,Vitelli2012,
Avalos2014,
Przedborski2015,Przedborski2015b}. Such systems are of broad interest primarily because of their usefulness for a variety of physical applications, ranging from shock mitigation~\cite{Nakagawa2003,Hong2005,Sokolow2005,Doney2006,Melo2006,Doney2009} 
to energy localization
~\cite{Vergara2006,Daraio2006,Breindel2011,Przedborski2015b}. Their usefulness for these applications stems from their ability to support the propagation of non-dispersive travelling disturbances such as SWs, which are a notable feature of many power-law interacting systems~\cite{Przedborski2015,Mohan2005,Valkering1980}. 

A propagating SW is initiated in an uncompressed chain by a simple edge impulse. However unlike solitons in continuum systems, SWs in these discrete systems suffer from weak interactions with each other~\cite{Job2005,Manciu2000,Manciu2002} and with system boundaries~\cite{Sen2004,Mohan2005,Sen2005,Job2005} since grains are capable of breaking contact. These interactions lead to SW-breakdown processes, the creation of secondary solitary waves (SSWs)~\cite{Manciu2000,Manciu2002,Job2005}, and subsequent energy exchanges~\cite{Avalos2007,Avalos2011} in SW-SW collisions.  

Long after singular perturbations to the chain, the system reaches an equilibrium-like, ergodic phase~\cite{Sen2004,Mohan2005,Sen2005,Avalos2007,Avalos2011,Avalos2014,Sen2008} marked by a large number of SSWs that are equally likely to be moving in either direction, called \emph{quasi-equilibrium} (QEQ). 
This phase exhibits unusually large~\cite{Sen2004,Mohan2005,Sen2005,Avalos2007,Avalos2011,Avalos2014,Sen2008} and occasionally persistent (rogue)~\cite{Han2014} fluctuations in system kinetic energy, which impedes energy equipartitioning among all grains in the system, making it distinct from true equilibrium. To the time scales previously considered in dynamical studies, QEQ was observed to be a general feature of systems with no sound propagation~\cite{Sen2004,Mohan2005,Avalos2011}. Until very recently, the question of whether QEQ is the final phase for such systems was a long-open problem. However, it has now been well-established~\cite{Onorato2015,Przedborski2017a,Przedborski2017b} that such systems with power-law interactions
can indeed achieve thermal equilibrium 
after sufficiently long times, and that the time scale to equilibrium increases with the degree of nonlinearity in the interaction potential. 

In homogeneous Hertzian chains, equilibrium was proved primarily by demonstrating energy equipartitioning among the \textit{independent degrees of freedom} in the system. This was accomplished by illustrating that the calculated finite heat capacity of the system agreed with the value predicted by Tolman's generalized equipartition theorem~\cite{Tolman1918}. Beyond this, it was shown that kinetic energy fluctuations relax to \emph{finite} values in finite interacting systems. Such values are influenced by the heat capacity in a $\bm{NVE}$ ensemble~\cite{Lebowitz1967,Rugh1998} and are governed by the exponent on the interaction potential. 

Hence, in finite systems in equilibrium, each grain does not have \emph{exactly} the same kinetic energy at any instant in time. Rather, each grain's kinetic energy fluctuates according to the same probability density function (PDF). In our previous work~\cite{Przedborski2017a} we derived approximations to the analytic form of the velocity and kinetic energy PDFs, different from hard spheres, and which incorporate the finite interaction potential, and these were found to agree well with results of particle dynamics simulations. In the current manuscript, we expand on our previous work to include heterogeneous systems, such as diatomic, where band gaps in the dispersion curve form, and tapered and random-mass chains, where many inertial mismatches leads to energy dispersion.  

The remainder of the paper is organized as follows. In Sec.~\ref{sec:Model} we introduce the model for the Hertzian chains and review the approximate PDFs for grain velocity and grain and system kinetic energies.  We also introduce the correlation functions used to monitor the onset of the transition to equilibrium. Then we give the details of the simulation parameters in Sec.~\ref{sec:methods}. In Sec.~\ref{sec:results}, we present the results, and finish with some concluding remarks in Sec.~\ref{sec:conclusions}.

\section{\label{sec:Model}Model and Theory}
The specific systems under consideration are heterogeneous 1D chains of $N$ grains, where each grain is characterized by mass $m_i$ and radius $R_i$. Adjacent grains interact via a Hertz-like contact-only potential~\cite{Hertz1882}. The Hamiltonian describing the system is:
\begin{equation}
\mathbf{H} = K+U = \frac{1}{2} \sum_{i=1}^N m_i \bm{v}_i^2 + \sum_{i=1}^{N-1} a_{i,i+1} \Delta_{i,i+1}^n,
\label{eq:hamiltonian}
\end{equation}
where $\bm{v}_i$ is the velocity of grain $i$ and $\Delta_{i,i+1} \equiv R_i + R_{i+1} - (x_{i+1}-x_i) \ge 0$ is the overlap between neighbouring grains, located at position $x_i$. If $\Delta_{i,i+1}<0$, there is no interaction. In the above expression, the exponent $n$ is shape-dependent ($n=2.5$ for spheres), and $a_{i,i+1}$ contains the material properties of the grains and the grain radii~\cite{Sun2011}. The grain interactions with the fixed walls adds two terms to the Hamiltonian~\cite{Przedborski2015}. 

In typical numerical simulations, the system is perturbed by giving an end grain an initial velocity directed into the chain at time $t=0$. This initiates the formation of a propagating SW, which eventually breaks down into a sea of secondary solitary waves (SSWs) after numerous collisions with boundaries. This breakdown process, which happens sufficiently long after the initial perturbation to the system, is facilitated by the formation of transient inter-grain gaps and can be modelled as a transition from a non-ergodic (SW) phase to an ergodic (equilibrium) phase. Since this late-time phase is characterized by a large number of SSWs traversing the system in either direction, energy is, on average, shared equally among all the grains. For systems with zero energy dissipation, a $\bm{NVE}$ ensemble is hence established. This means that the long-term dynamics of Hertzian chains is best described by the statistics of a 1D gas of interacting spheres in thermodynamic equilibrium.

\subsection{\label{sec:dist}$\bm{NVE}$ distribution functions}
It has been previously established that the PDF of particle velocity of a $d$-dimensional, finite sized $\bm{NVE}$ ensemble is not a Maxwell-Boltzmann distribution~\cite{Scalas2015,Ray1991}. Rather, the probability distribution across the phase space occupied by an $\bm{NVE}$ ensemble is:
\begin{equation}
\Omega_E = \frac{\delta\left(E-\mathbf{H}\right)}{\Omega},
\label{eq:joint_pdf}
\end{equation}
where $\delta(x)$ is the dirac delta function, and the normalization integral is found from the hypersurface defined by the shell with total energy $\mathbf{H}=E$ in a $2dN$-dimensional phase space,
\begin{equation}
\Omega = \int \delta\left(E-\mathbf{H}\right) \prod_{i=1}^N \prod_{\epsilon=1}^d dx_{i,\epsilon} dp_{i,\epsilon}.
\label{eq:ps_vol}
\end{equation}
The integral in equation~(\ref{eq:ps_vol}) is taken over all grain momenta $\bf{p}$ and all grain positions $\bf{x}$. For indistinguishable particles, multiplication of the integral by the pre-factor $1/(N!h^{dN})$ gives the classical density of states. Integration over the grain momenta is accomplished by scaling the momenta as $\tilde{p}_{i,\epsilon} = p_{i,\epsilon}/\sqrt{2m_i}$, and then introducing the spherical change of variable $\tilde{P}^2 = \sum_{i=1}^N\sum_{\epsilon=1}^d \tilde{p}_{i,\epsilon}^2$. Subsequent evaluation of the scaled momentum integrals gives the surface area of a $dN$-dimensional hypersphere of radius $(E-U)^{1/2}$, leaving the remaining integral over the grain positions:
\begin{displaymath}
\Omega = \frac{(2\pi)^{dN/2}}{\Gamma(dN/2)} \bigg( \prod_{i=1}^N m_i^{d/2} \bigg) 
 \times  \int \left (E-U \right )^{dN/2-1}\Theta\left(E-U\right) \prod_{i=1}^N \prod_{\epsilon=1}^d dx_{i,\epsilon},
\label{eq:dos}
\end{displaymath}
where $\Gamma(x)=(x-1)!$ is the Gamma function and $\Theta(x)$ is the Heaviside step function. 

An exact analytic solution for the Hamiltonian in equation~(\ref{eq:hamiltonian}) for a finite system with $U\neq0$ may be exceedingly difficult to derive. Thus, in our previous work~\cite{Przedborski2017a}, we approximated the integral over grain positions by using the virial theorem to replace $\left(E-U\right)$ with $\left(E-\langle U\rangle_v\right) = \langle K\rangle_v$, where $\langle \dots \rangle_v$ denotes the expected virial value, i.e. $\langle K \rangle_v = \frac{n}{n+2}E$, 
with $K$ the total \emph{system} kinetic energy. Thus the constant $\langle K\rangle_v$ can come out of the integral in equation~(\ref{eq:dos}), and the integral proceeds as previously described~\cite{Scalas2015,Ray1991,Shirts2006}. 

This substitution restricts the maximum momentum for each individual grain $i$ to a unique value based on its individual inertial mass, $|\bm{p}_i|_{\mathrm{max}} = \left ( 2m_i \langle K\rangle_v \right )^{1/2}$. Now the boundary of the momentum axes in phase space is set by the constant $\langle K\rangle_v$, and our analysis assumes that all states within this boundary are equally likely. However, in reality, the value of $\langle K \rangle_v$ is of course, an \emph{average} of the ensemble, and there are certainly grains with kinetic energy that, at times, are slightly greater than this value. Nevertheless, such fluctuations decrease with increasing $N$, guaranteeing that the number of phase space states beyond this limit is quite small, and we showed previously~\cite{Przedborski2017a} that the virial theorem value turns out to be a very good approximation for $N\gtrsim 10$.

The resulting PDF of per-grain velocities $v_i$ in 1D is then obtained by marginalization of the joint PDF, equation~(\ref{eq:joint_pdf}), giving~\cite{Scalas2015}
\begin{eqnarray}
\mathrm{PDF}(v_i) 
&=&\mathrm{B}\left(\alpha,\beta,\tilde{v}_i \right)/\left ( 2 \langle v_i \rangle_v \right), \nonumber \\
&=& \frac{1}{2\langle v_i \rangle_v} 
\left ( \frac{\Gamma(\alpha+\beta)}{\Gamma(\alpha)\Gamma(\beta)}
\left(\tilde{v}_i\right)^{\alpha-1}\left(1-\tilde{v}_i\right)^{\beta-1} \right ),
\label{eq:pdfv}
\end{eqnarray}
where
\begin{equation}
\tilde{v}_i = \frac{1}{2}\left(1+\frac{v_i}{\langle v_i \rangle_v}\right),
\end{equation}
with $\langle v_i \rangle_v^2=2\langle K\rangle_v/m_i$, and $\alpha=\beta=(N-1)/2$. In the above expression, $\mathrm{B}(\alpha,\beta,\tilde{v}_i)$ is the beta distribution, and $\Gamma$ is the gamma function. Since $\tilde{v}_i$ must lie in the interval $[0,1]$, it follows that $v_i \in [-\langle v_i\rangle_v, \langle v_i \rangle_v]$. Consequently, grains with different masses are characterized by different velocity distributions. Specifically, each velocity distribution is centred around $v_i=0$, but the width (variance) depends on the grain mass. 

In the limit $N\gg1$, equation~(\ref{eq:pdfv}) becomes the familiar Maxwell-Boltzmann 1D normal distribution,
\begin{equation}
\mathrm{PDF}(v_i) 
= {\cal N}\big(\mu,\sigma_i^2;v_i\big) = \frac{1}{\sigma_i \sqrt{2\pi}} \mathrm{e}^{-(v_i - \mu)^2/2\sigma_i^2}
\label{eq:MB}
\end{equation}
with mean $\mu=0$ and variance $\sigma_i^2=2\langle K\rangle_v/(N m_i)$. Here, ${\cal N}\big(\mu,\sigma_i^2)$ is the normal, or Gaussian, distribution.

While the variance of the distribution of grain velocities depends on grain mass, the distribution of kinetic energy per-grain $K_i$ is identical for each grain, regardless of its mass. The PDF of $K_i$ is obtained by making the replacement $v_i = \sqrt{2K_i/m_i}$ in $\mathrm{PDF}(v_i)$ and further employing the relation $\mathrm{PDF}(K_i)dK_i = 2\mathrm{PDF}(v_i) (dK_i/dv_i)^{-1} dK_i$, which gives the resulting beta distribution~\cite{Scalas2015,Shirts2006}
\begin{equation}
\mathrm{PDF}\left(K_i\right) = \mathrm{B}\left(\alpha,\beta; \tilde{K}\right)/\langle K\rangle_v,
\label{eq:pdfki}
\end{equation}
where $\tilde{K}=K_i/\langle K\rangle_v$, $\alpha=1/2$, and $\beta=(N-1)/2$. For $N\gg1$, this becomes the familiar Maxwell-Boltzmann distribution for kinetic energy, a gamma distribution $\mathrm{G}(\alpha,\beta,K_i)$:
\begin{equation}
\mathrm{PDF}\left(K_i\right) = \mathrm{G}(\alpha,\beta,K_i) = \frac{\beta^{\alpha}}{\Gamma(\alpha)}K_i^{\alpha-1}e^{-\beta K_i},
\label{eq:pdfki2}
\end{equation}
where $\alpha=1/2$ and $\beta=N/(2\langle K\rangle_v)$.
Both distributions predict an average kinetic energy per grain of $K_i = \langle K \rangle_v/N$, and a variance 
that is independent of grain mass. 

In our previous work, we derived an approximation to the distribution of system kinetic energy $K=\sum_{i=1}^N K_i$ from statistical theory by treating $K_i$ as independent and identically distributed (i.i.d.) variates drawn from the distribution of equation~(\ref{eq:pdfki2}). Using this method, the result is $\mathrm{PDF}\left(K\right) = \mathrm{G}(N/2,N/(2\langle K\rangle_v);K)$, which has the correct mean; however, the variance predicted by this distribution does not agree with the variance predicted by the finite system heat capacity. After trial-and-error, a better approximation was found to incorporate the exponent of the potential energy $n$,
\begin{equation}
\mathrm{PDF}\left(K\right) = \mathrm{G}\left(\frac{n+2}{2}\frac{N}{2},\frac{n+2}{2}\frac{N}{2\langle K\rangle_v}; K\right).
\label{eq:pdfk}
\end{equation}
This distribution not only gives an excellent match to the distribution calculated from molecular dynamics (MD) simulation~\cite{Przedborski2017a}, but it also has the correct variance as predicted by the equilibrium specific heat capacity in the $\bm{NVE}$ ensemble.

\subsection{\label{sec:cv}Specific heat}
An equilibrium value for the specific heat for Hertzian chains in the thermodynamic limit was derived previously~\cite{Przedborski2017a,Przedborski2017b} from an application of Tolman's generalized equipartition theorem~\cite{Tolman1918} to the Hamiltonian, equation~(\ref{eq:hamiltonian}). The result is
\begin{equation}
C_V^{\mathrm{Eq}} = \left ( \frac{n+2}{2n}\right ) k_B,
\label{eq:sp}
\end{equation}
which evidently depends \textit{only} upon the exponent in the potential, i.e. there is no dependence on grain (or wall) material, grain size, or temperature. equation~(\ref{eq:sp}) gives  the expected value of the specific heat in a $\bm{NVE}$ in the limit $N\gg 1$ when energy is equipartitioned among the degrees of freedom. 

This equilibrium specific heat also gives a prediction for the equilibrium fluctuations in total system kinetic energy, through the relation first derived by Lebowitz et al., which relates the two quantities in one-dimensional systems as~\cite{Lebowitz1967,Rugh1998} 
\begin{equation}
\frac{\langle\delta K^2 \rangle}{\langle K \rangle^2} = \frac{2}{N} \left( 1-\frac{1}{2 C_V}\right ), 
\label{eq:LPV}
\end{equation}
where $C_V$ is in units of $k_B$. When combined with equation~(\ref{eq:sp}), it follows that the expected variance in system kinetic energy is
\begin{equation}
\langle \delta K^2 \rangle = \frac{2}{N} \left( \frac{2}{n+2}\right) \langle K \rangle ^2, 
\label{eq:systemK_fluc}
\end{equation}
from which the factor of $(n+2)/2 < 1$ appears, which has already been included empirically as part of the distribution variance of equation~(\ref{eq:pdfk}). From equation~(\ref{eq:systemK_fluc}), it is clear that, in the equilibrium phase, $\langle \delta K^2 \rangle/\langle K \rangle ^2$ is absent of material dependence. This has been observed previously in MD simulations~\cite{Przedborski2015,Przedborski2017a,Przedborski2017b}.

Inverting equation~(\ref{eq:LPV}) provides one way to calculate the specific heat per grain from an MD simulation. Alternatively, one can use the exact formula for the microcanonical specific heat obtained by taking an energy derivative of the so-called microcanonical temperature, which in 1D gives~\cite{Rugh1998}
\begin{equation}
C_V = \frac{k_B}{N} \left ( 1 - \frac{(N-4) \langle 1/K^2 \rangle }{(N-2)\langle 1/K \rangle^2 } \right )^{-1}.
\label{eq:Cmc}
\end{equation}
In Sec.~\ref{sec:results}, we use both equations~(\ref{eq:LPV}) and~(\ref{eq:Cmc}) to calculate the specific heat from MD data for comparison with the predicted equation~(\ref{eq:sp}).

\subsection{Correlation functions}
The approximate form for the distribution of system kinetic energy was derived under the assumption of statistical independence between physical quantities, such as grain velocities. Here we introduce three grain correlation functions which we will later use with MD data to justify this assumption. 

We begin by looking for correlations in the time domain, and define the velocity auto-correlation function as:
\begin{equation}
C(t) = \sum_{i=1}^{N} C_i(t) = \sum_{i=1}^{N} \langle v_i(0) v_i(t) \rangle,
\label{eq:v_autocor_sum}
\end{equation}
where the angular brackets denote a convolution integral:
\begin{eqnarray}
C_i(t) &=& \left ( v_i(\tau) \ast v_i(-\tau) \right ) (t) \nonumber \\
&=& \int_{0}^{T} v_i(t+\tau) v_i(\tau) d\tau,
\label{eq:v_autocor}
\end{eqnarray}
with $T$ the length of the sampling interval. In practice, the integral in equation~(\ref{eq:v_autocor}) is typically computed using the convolution theorem, which gives $\left ( v_i(\tau) \ast v_i(-\tau) \right ) (t) = {\mathcal F}^{-1}\left ( {\mathcal F}(v_i)\cdot CC[{\mathcal F}(v_i)] \right )$, with $\mathcal F$ the fourier transform and ${\mathcal F}^{-1}$ its inverse, and $CC[\dots]$ denoting the complex conjugate. This function is used to indicate correlations in the time domain since any periodicity or history dependence in the grain velocity data will appear in the correlation function.

Similarly, we introduce a correlation function to quantify the amount of correlated motion in the spatial domain, i.e. between neighbouring grains. The neighbour momentum correlation function is defined by
\begin{equation}
p_{\textrm{c}}(t) = \sum_{i=1}^{N-1} p_{i}(t)p_{i+1}(t),
\label{eqn:cor}
\end{equation}
and the sign of the neighbour correlation function, 
\begin{equation}
\mathrm{sgn}\left(p_{\textrm{c}}(t)\right) = \frac{1}{N-1}
\sum_{i=1}^{N-1} 
\textrm{sgn}\left(p_{i}(t)\right)
\textrm{sgn}\left(p_{i+1}(t)\right)
\label{eqn:sgncor},
\end{equation}
where $p_i = m_i v_i$ denotes the momentum of grain $i$. Both these correlation functions give an estimate of the amount of correlated motion between neighbouring grains, which could result from SW propagation. (Note that we have chosen to monitor the correlations among grain momenta rather than grain velocities since we will be investigating heterogeneous chains comprised of grains with different masses.) 

In the non-ergodic SW phase, one expects there to be a large amount of correlated motion among the moving grains as the SW spanning several grains sweeps across the chain. In contrast, in the symmetric equilibrium phase, one expects there to be as many correlated as anti-correlated motions among neighbouring pairs of grains. Hence, we expect $p_{\textrm{c}}(t)$ to be non-zero early on, and drop to fluctuations about zero later on, indicating the onset of the equilibrium phase. Likewise, by only using the sign of the momentum in equation~(\ref{eqn:sgncor}), the actual number fraction of interfaces exhibiting correlated motion can be quantified, regardless of the amplitude of the momentum.

Since the grain momenta are not constant, and are rather described by a static probability distribution function in the equilibrium phase, it follows that $p_c(t)$ has an associated probability distribution function. When the grain momenta $p_i,p_{i+1}$ follow a normal distribution, which is the case for $N\gg1$, the exact analytic form for the distribution of neighbour correlations can be derived from statistical theory for homogeneous (i.e. single grain species) or diatomic (i.e. two species) chains, see the Appendix, with the result
\begin{equation}
\mathrm{PDF}(p_c)=\frac{|q|^{\frac{r-1}{2}}K_{\frac{r-1}{2}}\left(|q|\right)}
{\sigma_{p_1}\sigma_{p_2} 2^{\frac{r-1}{2}}\sqrt{\pi}\Gamma\left(\frac{r}{2}\right)},
\label{eqn:Q} 
\end{equation}
where $q=p_c/(\sigma_{p_1} \sigma_{p_2})$ (with $\sigma_{p_i}$ the standard deviation of the PDF of momentum for grain species $i$), $K_s(q)$ is a modified Bessel function of the second kind of order $s$, and $r=N-1$. Homogeneous chains correspond to $\sigma_{p_1}^2=\sigma_{p_2}^2 =2 m \langle K\rangle_v/N$, while diatomic chains have $\sigma_{p_1}^2= 2 m_1 \langle K \rangle_v/N \neq \sigma_{p_2}^2 = 2 m_2 \langle K \rangle_v/N$. Interestingly, equation~(\ref{eqn:Q}) gets wider with larger $N$ when $N\gg1$.

As a final measure of the correlations in Hertzian chains, we introduce the configurational temperature $T_c$, which for 1D systems is defined by~\cite{Jepps2000}
\begin{equation}
\frac{1}{k_B T_c} = \frac{\Big \langle - \sum_{i=1}^N \frac{\partial F_i}{\partial x_i} \Big \rangle}{\big \langle \sum_{i=1}^N F_i^2 \big \rangle},
\label{eq:Tc}
\end{equation}
where $k_B$ is Boltzmann's constant, and $F_i$ is the net force acting on particle $i$. The angular brackets denote an ensemble average, or equivalently, a time average in ergodic systems. This definition of temperature utilizes the configurational information contained within the particle interactions, rather than the kinetic information, to determine the temperature of the system.  In equilibrium for sufficiently dense systems, $T_c$ should equal the standard kinetic energy temperature~\cite{Jepps2000}.

We hypothesize that, neglecting the averaging in equation~(\ref{eq:Tc}) and monitoring the time dependence of the resulting rational quantity, the configurational temperature can act as a measure of the correlations and indicator to the onset of the transition to equilibrium in Hertzian chains. Particularly, early on in the erratic SW phase, the configurational temperature will exhibit large fluctuations, and will later relax to smaller fluctuations about a constant value as the system approaches the equilibrium phase. This relaxation should correspond to the onset of the transition to equilibrium in Hertzian chains.

\section{\label{sec:methods}Methods}
To examine the very long-time dynamics of Hertzian chains and closely inspect the transition to equilibrium, we ran MD simulations of various 1D configurations of $N$ grains held between fixed walls and described by the Hamiltonian in equation~(\ref{eq:hamiltonian}). These configurations include homogeneous (monatomic), diatomic, tapered, random radius, and random mass chains. 

In the monatomic chains, the grains are made of steel and 6~mm in radius, corresponding to a mass of 7075.4 mg. In diatomic chains, one of the species is comprised of steel grains 6~mm in radius, and the second grain species is comprised of grains whose Young's modulus and Poisson's ratio are equivalent to those of steel, but whose density has been altered to achieve the desired mass ratio. In the tapered chains, all grains are made of steel, and the largest grain is 6~mm in radius. A tapering percent (which controls the ratio of the radii of neighbouring grains) between 1-5\% is considered. Finally, in random radius and random mass chains, the Young's modulus and Poisson's ratio of all grains are equivalent to those of steel. For the random radius chains, the masses of all grains are kept constant at 7075.4 mg, while the radii are set by choosing random numbers between a fixed interval of 0.5-8 mm. Similarly, in the random mass chains, the grain radius is kept constant at 6 mm, while the masses are set with a random number generator within the range of 35-7075.4 mg.

Fixed walls comprised of steel are implemented in all systems, which adds two terms to the Hamiltonian as described in~\cite{Przedborski2015,Przedborski2017a,Przedborski2017b}. We do not apply any pre-compression, or squeezing of the chains, but rather each grain is initially just touching its neighbour between walls which are a distance of $\sum_{i=1}^N 2R_i$ apart. 

We consider values of the potential exponent $n$ from 2 (harmonic) to 4, and system sizes from $N=20$ to 100. A standard velocity Verlet algorithm is used to integrate the equations of motion with a 10~ps timestep, and no dissipation is included. The grains are set into motion with an asymmetric edge perturbation (initial velocity given to the first grain only, directed into the chain), causing a single initial SW to propagate through the system. The initial SW breaks down in collisions with boundaries and in the formation of gaps, creating numerous secondary solitary waves (SSWs). After a period of time, the number of SSWs increases to a point where the system enters into quasi-equilibrium~\cite{Sen2004,Mohan2005,Sen2005,Avalos2011,Avalos2014}. We allow the system to evolve for a substantial amount of time past this phase change. The system energy is constant to 10 significant digits for the entire simulation.

The time scale to QEQ onset is determined by the potential exponent $n$~\cite{Sen2008}, so we used the method described in reference~\cite{Przedborski2017b} to get an estimate for the optimal velocity perturbation for reaching equilibrium as quickly as possible. In most cases, it was necessary to collect at least one second of real time data, and even longer for larger values of $n$. Data of grain position and velocity are recorded to file every 10-100~$\mu$s, though we re-sample the data at time intervals beyond the dampening of velocity autocorrelation; typical sampling intervals were of the order of a few hundred $\mu$s. We call the last $20\%$ of each simulation the \emph{equilibrium interval}, and all further analysis is carried out with data from this interval. Here the deviation from the expected virial $\langle K \rangle_v = n/(n+2) E$ was $<1\%$ for all systems.

\section{\label{sec:results}Results and Discussion}
We begin by exploring several possible prerequisites to establishing equilibrium. There is no \textit{a priori} reason to assume the presence of ergodicity, or the absence of correlation or bias in systems with interaction potential energy exponent $n>2$, where energy can be transmitted via SWs whose width span several grains. In noisy data, such as the recorded velocity of a single grain, we must illustrate the validity of these common assumptions. 

Ergodicity is defined as the equivalence of ensemble and time averages of physical observables. It is thought that the QEQ phase in Hertzian systems is ergodic. One therefore might expect that the equilibrium phase is also ergodic, and we indeed establish this by a more rigorous statistical test than has been applied before. 

\begin{figure*}[htbp]
\centering
\includegraphics[width=0.88\textwidth]{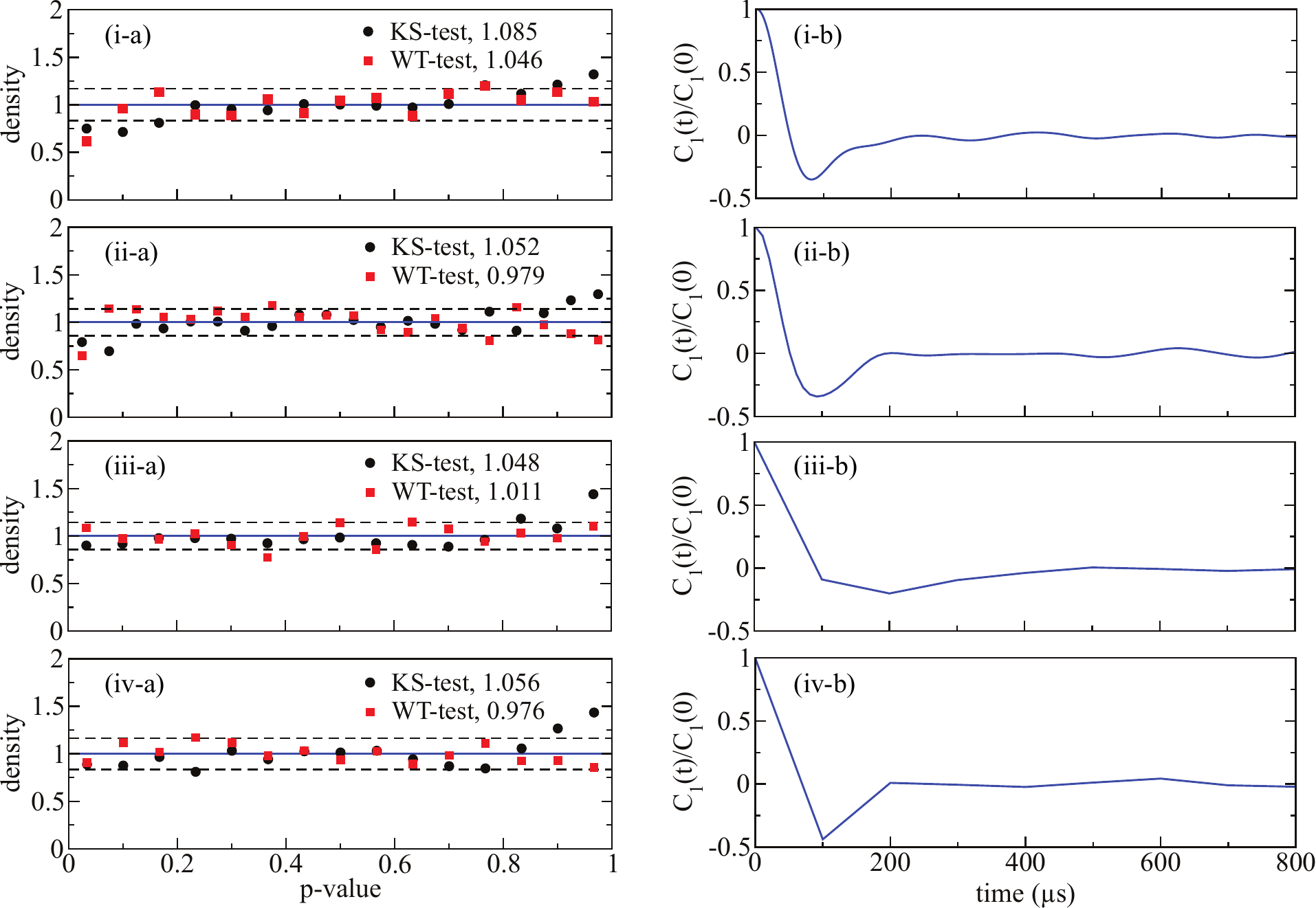}
\caption{(Color online) (a) Distribution of $p$-values for the two-sample KS- and WT-tests used to establish ergodicity in homogeneous Hertzian chains. Solid lines indicate the expected average value of the density, and dashed lines indicate the larger of the standard deviation, $\sigma_{KS/WT}$, of either test. The average densities of both tests are included in the top right corner of each plot.
(i) $n=2.0, N=100$; $\sigma_{KS}=0.167$, $\sigma_{WT}=0.145$; (ii) $n=2.5, N=38$; $\sigma_{KS}=0.134$, $\sigma_{WT}=0.138$;
(iii) $n=3.0, N=30$; $\sigma_{KS}=0.143, \sigma_{WT}=0.104$; (iv) $n=4.0, N=10$; $\sigma_{KS}=0.164, \sigma_{WT}=0.095$.
(b) Velocity autocorrelation function, equation~(\ref{eq:v_autocor}), for grain 1 computed over the entire equilibrium interval for various heterogeneous Hertzian chains. Data is re-sampled at time intervals where $C_1(t)/C_1(0)=0$ to ensure independence in the time domain. (i) $n=2.5$, $N=38$ homogeneous chain; (ii) $n=2.5$, $N=38$ diatomic chain with mass ratio $m_1/m_2 = 2$; (iii) $n=4$, $N=20$ tapered chain with tapering percent of 2.5\%; (iv) $n=3$, $N=20$ random-mass chain. Curves are not smooth in the last two plots because data was recorded to file at larger time intervals for these systems.}
\label{fig:fig1}
\end{figure*}

In homogeneous chains, under the null hypothesis, the time-domain velocity evolution of a single grain, and the velocities of the ensemble (i.e. all grains in the chain) at a given timepoint, should come from the same distribution. Furthermore, that distribution is nearly normally distributed. Thus to rigorously show ergodicity in homogeneous chains, we run repeated two-sample Kolmogorov-Smirnov tests (KS), and Welch's t-tests (WT), with both the single grain and ensemble time-point chosen at random. We plot a histogram of the distribution of 2500 $p$-values, shown in figures~\ref{fig:fig1}(i-a)-(iv-a) for four representative homogeneous systems. Under the null hypothesis of both tests, i.e. the system is ergodic, the $p$-values are uniformly distributed, which is the expected case if the underlying distribution is approximately normal~\cite{Bland2013}. The average densities for both tests, calculated as the weighted means of the distributions presented in figures~\ref{fig:fig1}(i-a)-(iv-a), are very close to one, as expected. There may be a minute upwards trend in the KS test, which might be indicative of slight skew in the underlying distribution. These effects are likely a result of the grain interactions with the confining walls. 

In comparison, in heterogeneous chains where the grains do not all have the same masses, the velocity distributions of each grain are not equivalent. Hence the velocities of the ensemble of grains (i.e. all grains in the chain) do not come from the same distribution as the time-domain velocity of any single grain. In other words, there is no equivalent grain which samples the same phase space as the ensemble comprised of all grains in the heterogeneous chain. While the velocity of each individual grain will be ergodic, as verified by statistical test in homogeneous chains, one cannot apply a statistical test simultaneously to all grains in the heterogenous chain to verify ergodicity. 

Since we aim to establish the absence of bias and correlations in the equilibrium phase in Hertzian chains, prior to further analysis below, we now remove bias in the time domain by computing the velocity autocorrelation function, equation~(\ref{eq:v_autocor_sum}), for each system. Since the grain velocities depend on grain mass, see equations~(\ref{eq:pdfv})--(\ref{eq:MB}), in heterogeneous systems it is more appropriate to consider the autocorrelation function of individual grains, equation~(\ref{eq:v_autocor}), rather than the sum over all grains. We present this correlation function for grain 1 (which is typically the largest and slowest moving grain) for four representative systems in figures~\ref{fig:fig1}(i-b)-(iv-b). We subsequently re-sample MD data for all systems at time intervals where velocity autocorrelation has vanished; typically of the order of a few hundred $\mu$s.

\begin{figure*}[htbp]
\centering
\includegraphics[width=0.88\textwidth]{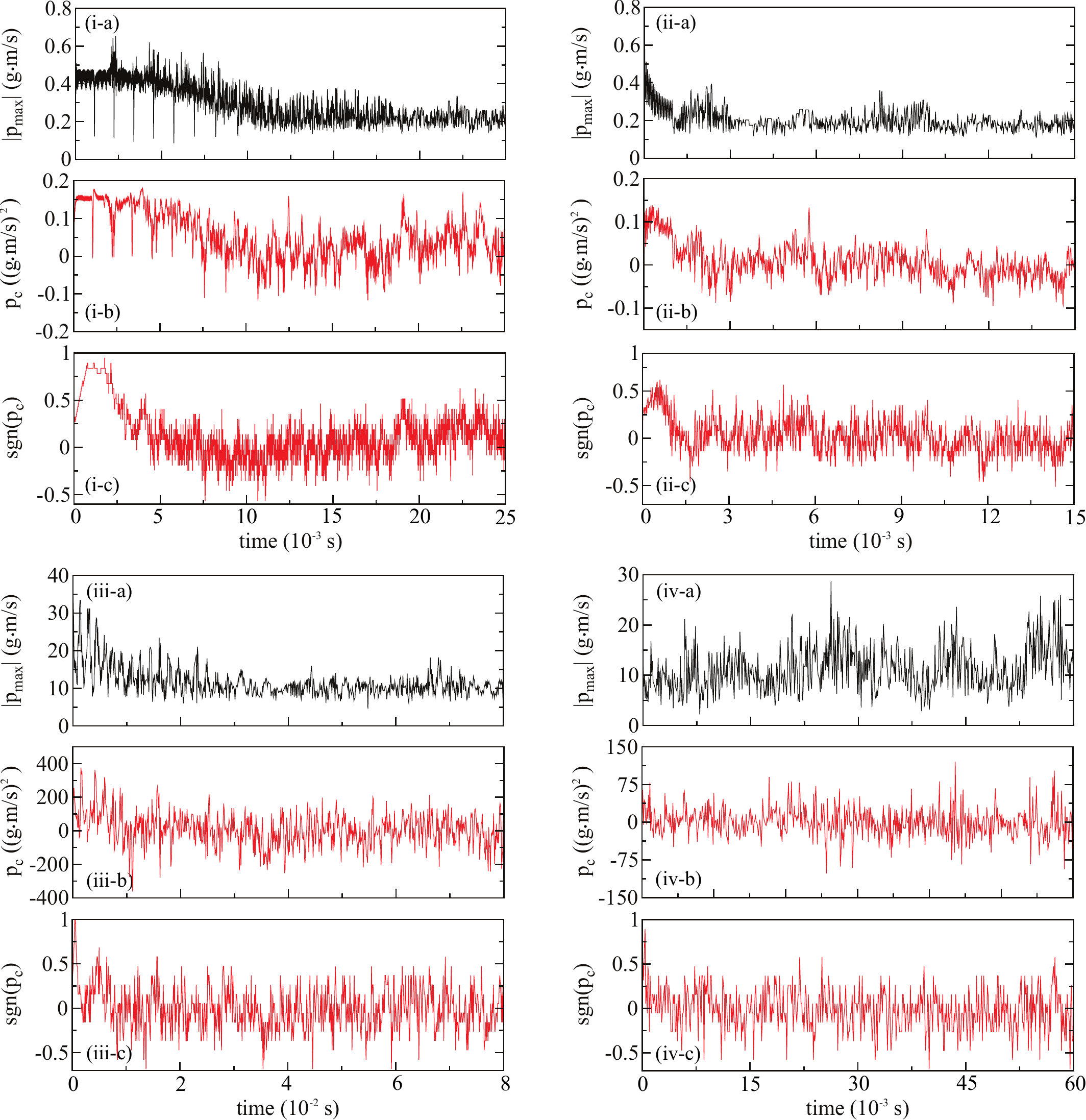}
\caption{(Color online) Grain-grain correlation functions for the four representative systems of figure~\ref{fig:fig1}(b). Plots labelled (i) correspond to the $n=2.5$, $N=38$ homogeneous chain, (ii) corresponds to the $n=2.5$, $N=38$ diatomic chain, (iii) to the $n=4$, $N=20$ tapered chain, and (iv) to the $n=3$, $N=20$ random mass chain. (a) labels the maximum absolute momentum of any grain in the chain as a function of time, (b) the neighbour momentum correlation as a function of time, equation~(\ref{eqn:cor}), and (c) the sign of the neighbour momentum correlation as a function of time, equation~(\ref{eqn:sgncor}). All quantities are computed from $t=0$.}
\label{fig:fig2}
\end{figure*}

While correlations in the time domain have been accounted for, we also test for correlated motion between neighbouring grains. Such correlations can be measured by monitoring various quantities, including the maximum absolute momentum of any grain in the chain, $|p_\mathrm{max}|$, as a function of time. This quantity is computed from $t=0$ for four representative systems in figures~\ref{fig:fig2}(i-a)-(iv-a). In homogeneous (monoatomic) chains, figure~\ref{fig:fig2}(i-a), $|p_\mathrm{max}|$ initially oscillates about a maximal value as the initial SW travels through the granular chain. The SW may make several passes through the chain before breaking down. Periodically, when the initial SW reaches a boundary and most of the system's energy converts to stored potential energy in the walls, $|p_\mathrm{max}|$ drops to a small value, before reflecting and resuming its course. Some time after the initial velocity perturbation, $|p_\mathrm{max}|$ relaxes to noisy oscillations about a much smaller value, denoting the onset of the QEQ phase. 

In diatomic chains, figure~\ref{fig:fig2}(ii-a), this relaxation happens much sooner due to the inertial mismatches between neighbouring grains. The SW breaks down much quicker since energy is both reflected and transmitted in successive collisions between neighbouring grains in such systems. In tapered chains, figure~\ref{fig:fig2}(iii-a), the progressive decrease in the radius and mass of the grains also causes the amplitude of the SW to decay more quickly than an equivalent homogeneous chain. This is reflected in the behaviour of $|p_\mathrm{max}|$, though it should be noted that in figure~\ref{fig:fig2}(iii-a), the time scale is longer since $n=4$ for this system, and the time scale to equilibrium increases with $n$. Finally, in random mass chains, figure~\ref{fig:fig2}(iv-a), $|p_\mathrm{max}|$ drops to noisy oscillations about a small value almost immediately, which insinuates that random mass chains do not support SW propagation. 

Interestingly, in random radius chains in which the grain masses are the same, the behaviour of $|p_\mathrm{max}|$ is closer to the behaviour exhibited by homogeneous chains. In particular, the relaxation is apparent since it happens over a longer timescale, implying that the random radius chains do support SW propagation. Similarly, in diatomic chains in which the grain masses are the same but the two grain species differ in their radii, the behaviour of $|p_\mathrm{max}|$ is nearly identical to that observed in the equivalent homogeneous chain. This indicates that it is the differences in grain inertia, and not the grain shape, that play a major role in the SW breakdown processes in these systems. 

To further monitor the grain correlations, we also calculate the neighbour momentum correlation function, equation~(\ref{eqn:cor}), from $t=0$ for all of our systems, and results are shown for the four representative systems in figures~\ref{fig:fig2}(i-b)-(iv-b). It is clear from these figures that the relaxation of $p_c$ identically mirrors that of $|p_\mathrm{max}|$, and that $p_c$ drops to oscillations about zero as the system enters into QEQ, as expected. We also present the results for the sign of the neighbour correlation function, equation~(\ref{eqn:sgncor}), computed from $t=0$ in figures~\ref{fig:fig2}(i-c)-(iv-c) for the same systems. Interestingly, $\mathrm{sgn}(p_c)$ has already dropped to zero roughly by the time that $p_c$ and $|p_\mathrm{max}|$ are beginning to relax.  

As evident from all the plots in figure~\ref{fig:fig2}, grain-grain correlations die out early in the simulation, indicating the onset of the equilibrium phase. We see that $\mathrm{sgn}\left(p_{\textrm{c}}\right)$ fluctuates between extremes of about $\pm0.5$, values that represent the difference in the number of interfaces between neighbouring grains whose motion are correlated (grains moving parallel) versus anti-correlated (grains moving anti-parallel). From these maxima, as much as 75\% of interfaces at times are either correlated or anti-correlated. The appearance of a large number of anti-correlated interfaces start very early in the simulation, and may trigger the onset of the equilibrium phase.

We conclude from the presence of equal amounts of correlated and anti-correlated grain motion that SSWs do not add correlated bias to the motions of neighbouring grains in the equilibrium phase. These results also confirm that the single-grain quantities can be treated as i.i.d. random variables drawn from the distributions of equations~(\ref{eq:pdfv}) or (\ref{eq:MB}) for grain velocity, and equations~(\ref{eq:pdfki}) or (\ref{eq:pdfki2}) for grain kinetic energy. 

\begin{figure}[htbp]
\centering
\includegraphics[width=0.45\textwidth]{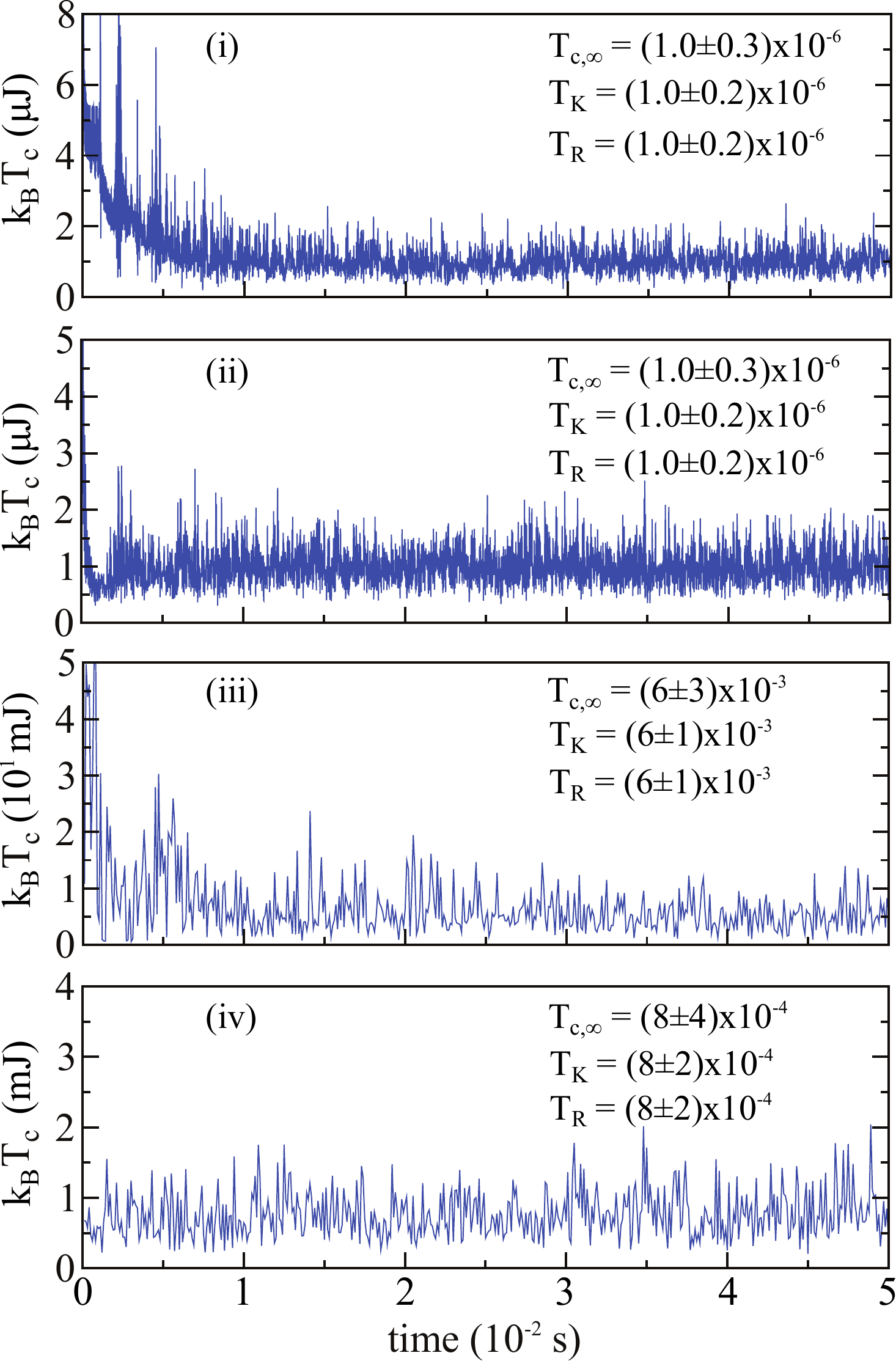}
\caption{(Color online) Configurational temperature, equation~(\ref{eq:Tc}) (with averaging neglected), for the four representative systems in figures~\ref{fig:fig1}(b) and~\ref{fig:fig2}. For plots (i)-(iii), the curves fit to an exponential function which decays to a nonzero constant value. This long-term constant value is denoted as $T_{c,\infty}$ and is presented on each plot (for the random mass chain in (iv), $T_{c,\infty}$ was obtained from a constant fit to the data). We also present the values of the kinetic temperature $T_K$ corresponding to kinetic energy $\langle K \rangle_v$, as well as the microcanonical temperature $T_R$ introduced by Rugh~\cite{Rugh1998} (see text for more details). Note that $T_{c,\infty}, T_K$, and $T_R$ are presented in units of $k_B$. (Plots (i)-(iii) are zoomed in to clearly show the size of the long-term fluctuations.)}
\label{fig:fig3}
\end{figure}

As a final probe of the correlations among grains, we compute the configurational temperature, equation~(\ref{eq:Tc}), for all of the systems. To monitor the convergence of $T_c$ over time, we drop the time averaging, and the results of the instantaneous ensemble temperature are shown for the four representative systems in figure~\ref{fig:fig3}. For the systems which support SW propagation, $T_c$ has an exponential decay to fluctuations about a constant value. Moreover, the behaviour of $T_c$ computed in this way mimics the behaviour of $|p_\mathrm{max}|$ and $p_c$. In particular, $T_c$ starts to settle to fluctuations about the long-term constant value at roughly the same time that $|p_\mathrm{max}|$ and $p_c$ have decayed, indicating the onset of equilibrium. Thus $T_c$ computed in this way also provides a reliable way to measure the SW breakdown rate in Hertzian chains.  

We see from figure~\ref{fig:fig3} that the long-term constant value of the configurational temperature, $T_{c,\infty}$, agrees with the kinetic temperature defined by $k_B T_K = 2\langle K \rangle_v/N$, as well as the microcanonical temperature defined by~\cite{Rugh1998} $1/(k_B T_R) = \big((N-2)/2\big)\langle 1/K \rangle$, within the error bars. The agreement is better when $N$ is large, see figures~\ref{fig:fig3}(i) and (ii). For smaller values of $N$, figures~\ref{fig:fig3}(iii) and (iv), the agreement is less since $T_c$ and $T_K$ are only accurate to $O(1/N)$. The drastic increase in the size of the error for the systems in figures~\ref{fig:fig3}(iii) and (iv) is a consequence of both the small system size and the large initial perturbation (a larger initial perturbation was given to these systems since $n$ is larger).    

\begin{figure*}[ht]
\centering
\includegraphics[width=0.98\textwidth]{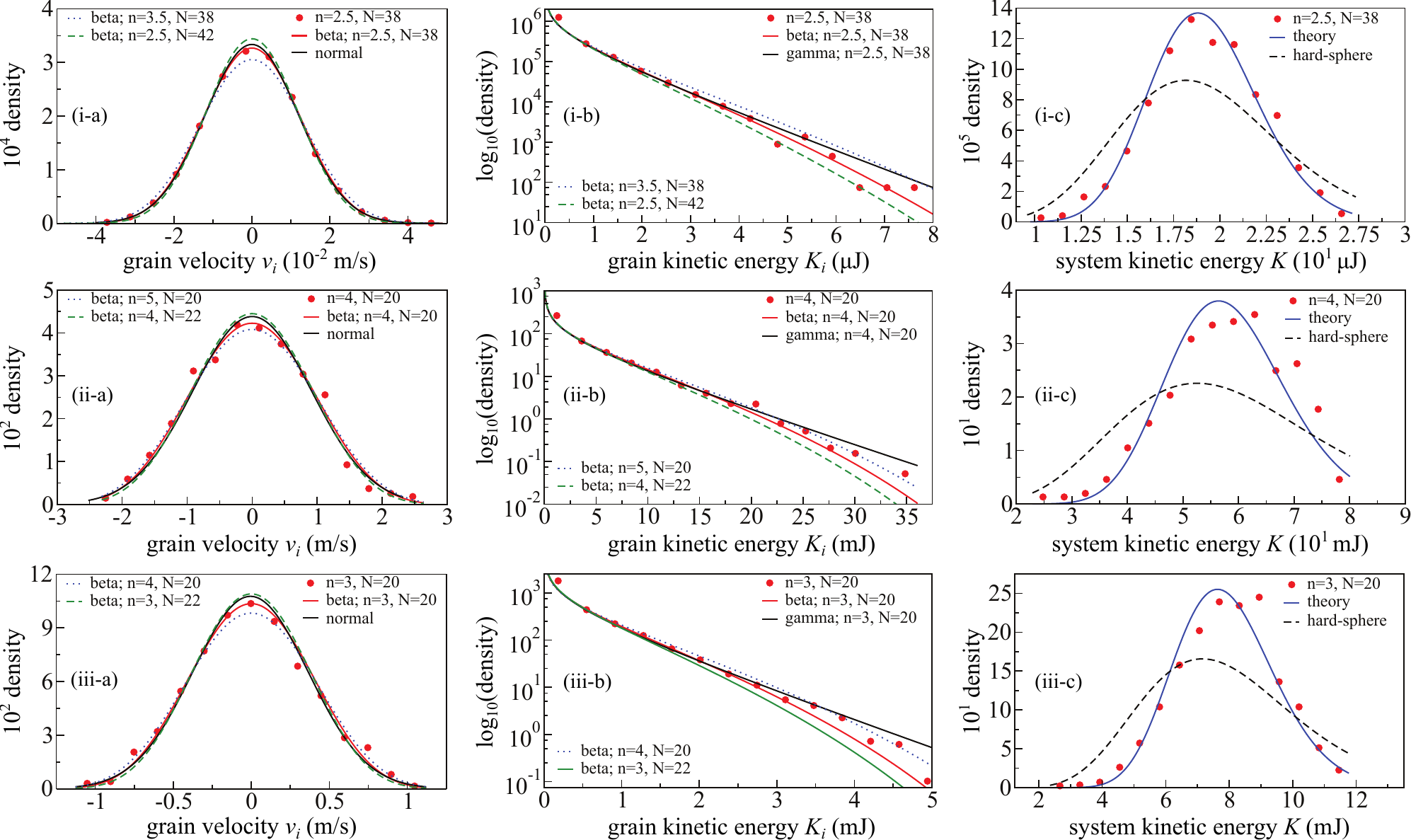}
\caption{(Color online) Distribution of grain velocity, grain kinetic energy, and system kinetic energy for the three heterogeneous systems in figures~\ref{fig:fig1}(b),~\ref{fig:fig2}, and~\ref{fig:fig3}. Results of MD simulations are shown as filled circles. In columns (a) and (b), solid lines are predicted distributions (equations~(\ref{eq:pdfv}),~(\ref{eq:MB}),~(\ref{eq:pdfki}), and ~(\ref{eq:pdfki2})), and dashed/dotted lines are the corresponding distributions with parameters slightly changed to illustrate the sensitivity of equations~(\ref{eq:pdfv}) and~(\ref{eq:pdfki}). In column (c), solid curve is the theoretical prediction equation~(\ref{eq:pdfk}), and dashed line is the corresponding hard-sphere distribution.}
\label{fig:fig4}
\end{figure*}

Now that we have demonstrated the absence of correlations in the long-term phase of Hertzian chains, we show that this phase is indeed an equilibrium phase. To accomplish this, we check the distributions of grain velocities and kinetic energy, as well as the equipartitioning of energy among all grains via the specific heat. First we test the grain distribution functions presented in Sec.~\ref{sec:dist}, and show the agreement between the expected PDFs (equations~(\ref{eq:pdfv}),~(\ref{eq:MB})--(\ref{eq:pdfk})) and MD data for three representative heterogeneous systems in figure~\ref{fig:fig4}. In each system, the per-grain velocity data agrees with the beta distribution, equation~(\ref{eq:pdfv}), which is nearly identical to the normal distribution, equation~(\ref{eq:MB}), for large $N$, see figures~\ref{fig:fig4}(i-a)-(iii-a). 

The grain kinetic energy distributions are presented in figures~\ref{fig:fig3}(i-b)-(iii-b), illustrating agreement between MD results and equation~(\ref{eq:pdfki}) for large $N$. The difference between equations~(\ref{eq:pdfki}) and~(\ref{eq:pdfki2}) looks fairly pronounced in the log scale with smaller $N$, where the beta distribution generally has a cutoff before the tail of the MD data. The area under the MD histogram past the beta-distribution cutoff at $\langle K \rangle_v$ illustrates the earlier point of the small number of states beyond the limits of phase space used to derive equation~(\ref{eq:pdfki}). However, this area is exaggerated in the log scale plots; it was found previously~\cite{Przedborski2017a} that for homogeneous systems with $N=10$, $P(K_i>\langle K \rangle_v)\lesssim 0.05\%$, while for larger $N$ it's even less. 

The sensitivity of equations~(\ref{eq:pdfv}) and~(\ref{eq:pdfki}) to $n$ and $N$ are also shown in figures~\ref{fig:fig4}(i-a)--(iii-a) and (i-b)--(iii-b), by plotting curves of incorrect values of $n+1$ or $1.1N$. They do not agree as well with the data, and illustrate that the predicted distributions are indeed the best fit to the data.

figures~\ref{fig:fig4}(i-c)--(iii-c) contain the distributions of system kinetic energy from MD simulations, along with the corresponding equation~(\ref{eq:pdfk}), for the three representative heterogeneous systems. The agreement between MD data and the expected result is very good for $N=38$, see figure~\ref{fig:fig4}(i-c); however there is a slight skew in equation~(\ref{eq:pdfk}) when $N=20$, figures~\ref{fig:fig4}(ii-c) and (iii-c), which was also observed in homogeneous systems~\cite{Przedborski2017a}. For comparison, we also present the distribution without the variance correction, $\mathrm{G}(N/2,N/(2\langle K\rangle_v);K)$, which we call the hard-sphere limit, and clearly does not agree with any MD data of interacting grains. 

\begin{table}[htbp]
\caption{\label{table1}Specific heat capacity calculated from MD simulation data for various heterogeneous chains using equation~(\ref{eq:Cmc}) and (the inverted)  equation~(\ref{eq:LPV}), and the expected equilibrium value in the thermodynamic limit, equation~(\ref{eq:sp}).}
\begin{indented}
\item[]\begin{tabular}{p{1.5cm}p{1.5cm}p{2.0cm}p{2.0cm}p{2.0cm}}
\br
$n$ & 
$N$ & 
$C_V^\mathrm{Eq}/k_B$ (\ref{eq:sp})& 
$C_V/k_B$ (\ref{eq:LPV})& 
$C_V/k_B$ (\ref{eq:Cmc})\\   
\mr
\multicolumn{5}{c}{Diatomic chains} \\
2.5 & 38 & 0.900 & 0.887 & 0.870 \\
2.5 & 100 & 0.900 & 0.910 & 0.896 \\	                      
3 & 20 & 0.833 & 0.797 &  0.846 \\	
3 & 50 & 0.833 & 0.834 & 0.843 \\
4 & 20 & 0.750 & 0.751 & 0.753 \\
4 & 100 & 0.750 & 0.748 & 0.731 \\
\mr
\multicolumn{5}{c}{Tapered chains} \\
2.5 & 20 & 0.900 & 0.908 & 0.863 \\
2.5 & 50 & 0.900 & 0.914 & 0.909 \\
3 & 50 & 0.833 & 0.831 & 0.809\\
4 & 20 & 0.750 & 0.749 & 0.747 \\
\mr
\multicolumn{5}{c}{Random mass chains} \\
2.5 & 20 & 0.900 & 0.909 & 0.908 \\	
2.5 & 38 & 0.900 & 0.898 & 0.901 \\
3 & 20	& 0.833 & 0.811 & 0.828 \\
3.25 & 15 & 0.808 & 0.803 & 0.806 \\
3.5 & 25 & 0.786 & 0.797 & 0.787 \\
\br  
\end{tabular}
\end{indented}
\end{table}

Lastly, we compute the specific heats of MD simulation data using both equations~(\ref{eq:LPV}) and~(\ref{eq:Cmc}) to address the issue of equipartitioning of energy in these systems. These calculated results are directly compared with $C_V^\mathrm{Eq}$ predicted by equation~(\ref{eq:sp}) in Table~\ref{table1} for various heterogeneous chains. It is evident that  for larger $N$, the values calculated by equations~(\ref{eq:LPV}) and~(\ref{eq:Cmc}) agree very well with the theory. Moreover, even for small ($N\lesssim20$) systems, the deviation from theory is no more than $\sim 5\%$, and improve with additional data points in the averaging. 

The fact that the calculated specific heat agrees with the value predicted by the generalized equipartition theorem for $N\gg1$ provides evidence that energy is indeed equipartitioned in the heterogeneous Hertz chain at long enough times. This establishes that the very long-time dynamics of 1D heterogeneous granular chains with zero dissipation is a true equilibrium phase~\cite{Avalos2011}. 

While we have presented results for heterogeneous chains given asymmetric perturbations, it should be noted that these results are unchanged when the systems are given symmetric edge perturbations. The exception to this is when the symmetric edge perturbations induce a mirror reflection symmetry about the centre of the chain, such as in a homogeneous chain or an odd-$N$ diatomic chain. This symmetry results in a loss of degrees of freedom in the system, as discussed in Refs.~\cite{Przedborski2017a} and \cite{Przedborski2017b}, and the microcanonical specific heat and the PDF of system kinetic energy must be modified to account for this. In this case, it is important to stress that energy is equipartitioned \textit{among the independent degrees of freedom} in the system at long times. In systems in which there is no mirror reflection symmetry, the number of independent degrees of freedom and the number of grains are equivalent, thus such a distinction is not required.

\section{\label{sec:conclusions}Conclusions}  
We have illustrated that the long-term dynamics of 1D granular systems between fixed walls and with zero dissipation is a true equilibrium phase~\cite{Avalos2011}. In particular, we first used statistical tests to rigorously establish that the long-term dynamics is ergodic. Then we monitored correlations among grains via the neighbour momentum correlation functions and the configurational temperature. We showed that that correlations among grains vanish early on, indicating the onset of the transition to equilibrium.  

Moreover, we expanded on our previous work~\cite{Przedborski2017a,Przedborski2017b} 
to include heterogeneous chains, and showed that grains of different masses are characterized by different velocity distributions. We also showed that the approximate distribution functions for grain velocity, grain kinetic energy, and system kinetic energy that were derived previously for interacting particles in a microcanonical ensemble~\cite{Przedborski2017a,Przedborski2017b} agree well with MD data for various heterogeneous systems, including diatomic, tapered, and random mass chains. Lastly, we illustrated that energy is equipartitioned at long times in these systems by showing agreement between calculated specific heat capacities from MD data and expected equilibrium values.   

Most interestingly, we provided evidence that, apart from the degree of nonlinearity in the system, the configuration of masses influences the timescale of the transition to equilibrium. In particular, the transition can be accelerated by introducing inertial mismatches between grains. This is best demonstrated by the random mass chains, which do not support SW propagation and are therefore seen to start to equilibrate much sooner than homogeneous chains with the same degree of nonlinearity in the contact potential. This result may be useful for physical applications such as shock disintegration. 

It would be interesting to see how these ideas extend to systems with driving and dissipation. 

\appendix
\section*{Appendix}\label{sec:App_A}
\setcounter{section}{1}

To obtain an analytic expression for $\mathrm{PDF}(p_c)$, assuming the underlying distribution of particle velocities is given by~equation~(\ref{eq:MB}), it is easiest to proceed in two steps. First, we determine the distribution function for the product of two neighbouring grain momenta. Then we determine the distribution function for the sum of such products. 

In the limit $N\gg1$, equation~(\ref{eq:pdfv}) (and equation~(\ref{eq:MB})) predicts $\sigma_i^2 \equiv \mathrm{var}(v_i) = 2\langle K \rangle_v/(Nm_i)$, which immediately reveals that $\sigma_{p_i}^2 \equiv \mathrm{var}(p_i) = 2 m_i \langle K \rangle_v/N$. We assume the grain momenta can be treated as i.i.d random variates drawn on a normal distribution with zero mean and variance $\sigma_{p_i}^2$, i.e. $p_i = X \sim\mathcal{N}(0,\sigma_{p_1}^2)$ and $p_{i+1} = Y \sim\mathcal{N}(0,\sigma_{p_2}^2)$. These represent the momenta of the even and odd numbered grains. We are then first interested in the distribution of the product $Z\equiv XY$.

To compute the distribution of $Z$, we consider the characteristic function of the distribution of $X$ (or equivalently of $Y$). For a scalar random variable $X$, the characteristic function $\varphi_X(t)$ is defined as the expected value of $\exp(itX)$:
\begin{eqnarray}
\varphi_X(t) &=& {\bm E}\left ( e^{itX} \right ) \equiv \int_{-\infty}^{\infty} e^{itx} f_X(x) dx,
\label{eq:charfnc}          
\end{eqnarray}
where $i=\sqrt{-1}$, $t\in \mathbb{R}$, and $f_X(x)$ is the probability density function. From equation~(\ref{eq:charfnc}), it is evident that $\varphi_X(t)$ is simply the inverse Fourier transform of $f_X(x)$, where $f_X(x) = \mathcal{N}(0,\sigma_{p_1}^2)$ in this case. The inverse Fourier transform of a Gaussian is a well-known result, thus for the momentum distributions we have  $\varphi_X(t) = \exp(-\sigma_{p_1}^2t^2/2)$. 

Using the law of total expectation~\cite{Weiss2005}, it follows that ${\bm E}(X) = {\bm E} ( {\bm E} (X|Y) )$, where ${\bm E}(X|Y)$ denotes the conditional expectation value, i.e. the expected value of $X$ given that $Y=y$. Now ${\bm E}\left ( e^{itXY} \right | Y=y )$ is simply $\exp(-\sigma_{p_1}^2t^2Y^2/2)$, and from the law of total expectation it follows that the characteristic function of $Z$, $\varphi_{Z}(t) \equiv \varphi_{XY}(t) = {\bm E}\left( e^{itXY} \right )$, is given by:
\begin{eqnarray}
 \varphi_{Z}(t) &=& {\bm E} \left (  {\bm E} \left ( e^{itXY} | Y=y \right ) \right ) \nonumber \\
&=& {\bm E} \left ( \exp(-\sigma_{p_1}^2t^2Y^2/2)  \right ) \nonumber \\
&=& \frac{1}{\sqrt{2\pi}\sigma_{p_2}}\int_{-\infty}^{\infty} e^{-(\sigma_{p_1} ty)^2/2} e^{-y^2/(2\sigma_{p_2}^2)}  dy \nonumber \\
&=& \frac{1}{\sqrt{1+ \sigma_{p_1}^2\sigma_{p_2}^2t^2}}.
\label{eq:char_z}
\end{eqnarray}
We obtain $\mathrm{PDF}(Z)$ by inverting its characteristic function, which is equivalent to taking the Fourier transform of $\varphi_Z(t)$:
\begin{eqnarray}
\mathrm{PDF}(Z) &=& \frac{1}{2\pi} \int_{-\infty}^{\infty} e^{-izt} \varphi_Z(t) dt,
 \nonumber \\
&=& \frac{1}{\pi\sigma_{p_1}\sigma_{p_2}} K_0\left(\frac{|z|}{\sigma_{p_1}\sigma_{p_2}}\right),
\label{eqn:Z}
\end{eqnarray}
where $K_0(z)$ is the modified Bessel function 
of second kind and order zero, and equation~(\ref{eq:char_z}) was used in obtaining the final expression. equation~(\ref{eqn:Z}) is the well-known product normal distribution.

The neighbour momentum correlation function in equation~(\ref{eqn:cor}) involves the sum of terms like $Z$, so we now let $Q=\sum_{j=1}^{r} Z_j$ be the sum of $N=r+1$ independent variates $Z_j$ drawn from $\mathrm{PDF}(Z)$. By definition, the characteristic function of $Q$ is
\begin{eqnarray}
\varphi_Q(t) &=& {\bm E} \left ( e^{itQ} \right ) = {\bm E} \left ( e^{it\sum_{j=1}^r Z_j} \right ) \nonumber \\
&=& \int_{\mathbb{R}^r} \bigg( \prod_{j=1}^r e^{it z_j} \bigg) f_{Z_1,\dots,Z_r}(z_1,\dots,z_r) dz^r,
\label{eq:char_Q}
\end{eqnarray}
where $f_{Z_1,\dots,Z_r}(z_1,\dots,z_r)$ is the joint probability density function of all $Z_j$, i.e. $P(Z_1=z_1, \dots, Z_r=z_r)$, and $dz^r$ denotes the product $dz_1\dots dz_r$. The integral is taken over the $r$-dimensional real space $\mathbb{R}^r$. The statistical independence of $Z_j$ implies that $f_{Z_1,\dots,Z_r}(z_1,\dots,z_r) = \Pi_{j=1}^{r}f_{Z_j}(z_j)$, thus from equation~(\ref{eq:char_Q}),
\begin{eqnarray}
\varphi_Q(t) &=& \int_{\mathbb{R}^r}  \bigg( \prod_{j=1}^r e^{it z_j} f_{Z_j}(z_j) \bigg) dz^r \nonumber \\
&=& \bigg ( \int_{-\infty}^{\infty} e^{it z} f_{Z}(z) dz \bigg )^r \nonumber \\
&=& \left ( 1 + \sigma_{p_1}^2\sigma_{p_2}^2 t^2 \right )^{-\frac{r}{2}},
\end{eqnarray}
where equation~(\ref{eq:char_z}) was used in obtaining the last expression. Finally, the distribution function for $Q$, i.e. $\mathrm{PDF}(p_c)$, is obtained from a Fourier transform of this last equation. 

\begin{figure*}[ht]
\centering
\includegraphics[width=0.88\textwidth]{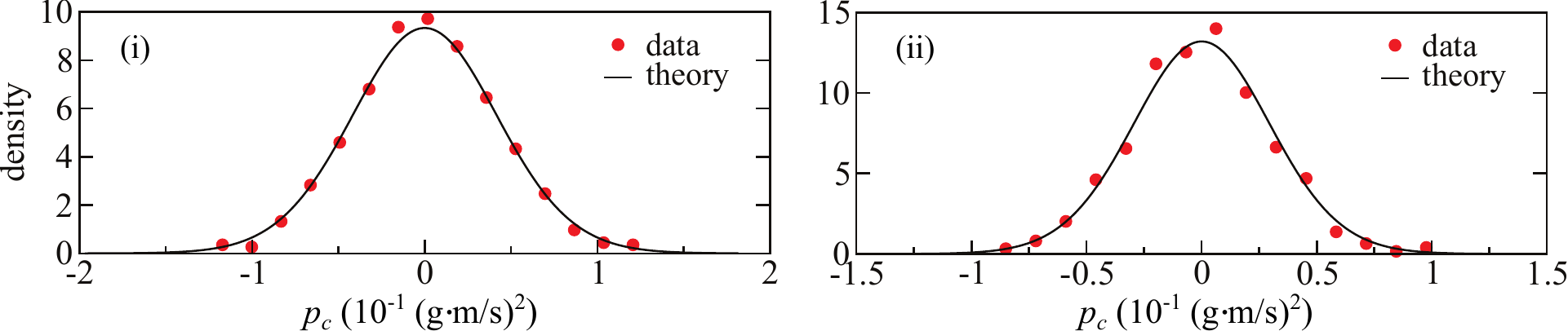}
\caption{(Color online) Distribution of the neighbour momentum correlation $p_c$ within the equilibrium phase. Circles correspond to MD data, and solid line to the predicted curve, equation~(\ref{eqn:Q}). (i) corresponds to a homogeneous $n=2.5, N=38$ system, and (ii) to a diatomic $n=2.5, N=38$ system.}
\label{fig:fig5}
\end{figure*}

We show agreement between the predicted $\mathrm{PDF}(p_c)$ and MD data for a homogeneous chain ($\sigma_{p_1}^2 = \sigma_{p_2}^2 = 2m\langle K\rangle_v/N$) in figure~\ref{fig:fig5}(i) and for a diatomic chain ($\sigma_{p_1}^2 = 2m_1\langle K\rangle_v/N \neq \sigma_{p_2}^2 = 2m_2\langle K\rangle_v/N$) in figure~\ref{fig:fig5}(ii). Symmetric and centred at zero, it is clear that the data agrees very well with equation~(\ref{eqn:Q}).

\ack This work was made possible by the facilities of the Shared Hierarchical Academic Research Computing Network (SHARCNET) and Compute/Calcul Canada.

\section*{References}
\bibliographystyle{iopart-num}
\bibliography{manuscript}

\end{document}